\documentclass[useAMS,usenatbib,mnras]{mn2e}
\usepackage{aas_macros}
\usepackage[pdftex]{graphicx}
\usepackage{amsmath}
\def\vela{PSR~J0835$-$4510}

\title[Single Pulse Timing]{Improving Pulsar Timing Precision with Single Pulses}

\author[M. Kerr]
{ M. Kerr$^{1}$\thanks{E-mail: matthew.kerr@gmail.com}\\
$^{1}$CSIRO Astronomy and Space Science, Australia Telescope National
Facility, PO~Box~76, Epping NSW~1710, Australia}
\begin{document}

\date{Accepted 2014 May 15. Received 2014 May 01; in original form
2014 April 30}


\maketitle

\label{firstpage}

\begin{abstract}
The measurement error of pulse times of arrival (TOAs) in the high S/N
limit is dominated by the quasi-random variation of a pulsar's
emission profile from rotation to rotation.  Like measurement noise,
this noise is only reduced as the square root of observing time,
posing a major challenge to future pulsar timing campaigns with large
aperture telescopes, e.g. the Five-hundred-metre Aperture Spherical
Telescope and the Square Kilometre Array.

We propose a new method of pulsar timing that attempts to approximate
the pulse-to-pulse variability with a small family of `basis'
pulses.  If pulsar data are integrated over many rotations, this basis
can be used to measure sub-pulse structure.  Or, if high-time
resolution data are available, the basis can be used to `tag' single
pulses and produce an optimal timing template.  With realistic
simulations, we show that these applications can dramatically reduce
the effect of pulse-to-pulse variability on TOAs.
Using high-time resolution data taken from the bright \vela{} (Vela),
we demonstrate a 25--40\% improvement in TOA precision.  Crucially for
pulsar timing applications, we further establish that these techniques
produce TOAs with gaussian residuals.

Improvements of this level halve the telescope time required to reach
a desired TOA precision.  Although some gains can be achieved with
existing data, the greatest improvements result from the `tagging'
approach, which in turn requires online or posthoc analysis of single
pulses, an important consideration for the design of future
instrumentation.

\end{abstract}


\section{Introduction}

From the discovery of the very first pulsar by \citet{Hewish68}, two
key features of pulsar emission were evident: (1) the extraordinary
regularity of the pulse train and (2) the wide variety of the shapes
and intensities of the single pulses of the train.  The former admits
a number of important experiments via pulsar timing.  E.g., the timing
of unrecycled and mildly recycled pulsars provided the first evidence
of gravitational radiation \citep{Taylor79} and other stringent tests
of general relativity \citep{Kramer09,Ransom14}, as well as the
discovery of magnetospheric reconfiguration
\citep{Kramer06,Lyne10,Hermsen13}.  Timing millisecond pulsars (MSPs)
probes physics at supra-nuclear density via neutron star mass
measurements \citep{Lattimer04,Demorest10,Antoniadis13} and enables
searches for low frequency gravitational waves
\citep[e.g.][]{Yardley11,Demorest13,Shannon13}.

The precision of timing experiments is limited by a variety of
stochastic noise phenomena.  The effects of propagation through the
interstellar medium are strong but are largely correctable with
broadband observations; see e.g. \citet{Keith13}.  For truly faint
pulsars, the limiting factor is the white radiometer noise added to
the observed pulse profile.  For most pulsars, however, the primary
limit arises from `timing noise', a poorly-understood collection of
phenomena in which the residuals of pulse times of arrival (TOAs) show
time-correlated structure, i.e. red noise.  Because timing noise seems
to be correlated with spin-down luminosity and other proxies for
pulsar age \citep[e.g.][]{Hobbs10b}, it most strongly affects
unrecycled pulsars.  However, its presence has also been detected in
MSPs \citep{Kaspi94,Manchester13}, where it dilutes sensitivity to the
correlated residuals expected to be induced by gravitational waves.

A second phenomenon---which in the literature has been called both
`jitter noise' \citep{Shannon10} and `stochastic wide-band impulse
modulated self-noise' \citep[SWIMS;][]{Oslowski11}---arises from the
second fundamental property of pulsar emission, its pulse-to-pulse
variability.  Despite the formidable nomenclature, the phenomenon is
simple: because we only record a finite number of pulses in an
observation, the resulting profile is randomly distorted from the
mean, biasing the measurement of phase.

Like radiometer noise, which fundamentally limits any radio
experiment, jitter/SWIMS seems to primarily induce white noise in
timing residuals and, for a given observation, decreases as $\propto
1/\sqrt{t}$ \citep{Shannon14}.  Thus, the ratio of jitter/SWIMS to
radiometer noise, $\sigma_j/\sigma_r$, depends only on the properties
of the pulsar and the receiving system, and historically, the
jitter/SWIMS phenomemon has only been important for the brightest
pulsars.  But increasingly sensitive receiver systems have brought the
issue to the forefront and several studies have been published in
recent years.  In particular, \citet{Oslowski11} present both an
excellent overview of the phenomenon (Dear Interested Reader: run, do
not walk, to print out a copy of this excellent paper) and a study of
its effect on PSR~J0437$-$4715, by two orders of magnitude the
brightest MSP.  Those authors conclude jitter/SWIMS dominates
measurement uncertainty, contributing at about thrice the level of
radiometer noise.

Because PSR~J0437$-$4715 is so bright, it is a bellwether for future
timing experiments with large aperture telescopes, e.g. FAST (the
Five-hundred-metre Aperture Spherical Telescope) and the SKA (Square
Kilometre Array).  Many pulsars whose timing is currently limited by
telescope sensitivity will be limited by jitter/SWIMS, with severe
implications for high precision timing experiments
\citep[e.g.][]{Hobbs14}.  In particular, without the ability to
mitigate jitter/SWIMS, the only tractable solution will be to time
large numbers of MSPs at lower precision.  This, in turn, places
substantial pressure on the design of FAST and SKA, viz. slew time for
the former and the availability of many tied subarrays for the latter.

In this work, we consider a new method for treating jitter/SWIMS based
on analysis of single pulses.  By approximating the pulse-to-pulse
variability as arising from a finite family of pulse shapes, we
attempt to estimate the particular realization of sub-pulse structure
in a given observation and from this produce a more accurate TOA.  We
give an overview of this approach in the context of pulsar timing in
\S\ref{sec:timing}.  In the following \S\ref{sec:observations}, we
briefly describe high-time resolution observations of the Vela pulsar
(\vela{})  before going on to discuss the approximation of its
pulse-to-pulse variabilty by construction of a template basis in
\S\ref{sec:basis}.  In \S\ref{sec:prof} and \S\ref{sec:tag}, we
discuss application of the basis method to simulated and real data
collected both with and without single pulses, respectively, and
demonstrate substantial improvement in TOA precision in both cases.
Finally, we summarize our results and discuss their implication for
and possible implementation in future timing experimets.

\section{Pulsar Timing}
\label{sec:timing}

Elementary pulsar timing comprises a series of measurements of the
time of arrival of a pulse in the laboratory frame.  Rather than
working directly with time series, it is typical to work instead in
terms of pulsar phase and to measure the phase offset, $\delta$, between
an observed pulse shape, $p$, and a template pulse shape, $t$.  If
pulse shape and template are realized as $N_p$-bin vectors
\textbf{p} and \textbf{t}, then the
probability density function (pdf) for \textbf{p} is given
by
\begin{equation}
\label{eq:basic_like}
f(\mathbf{p}\,|\,\delta,s,\mathbf{t},\mathbf{\sigma}) =
\prod_{i=1}^{N_p}\frac{1}{\sqrt{2\pi}\sigma_i}\exp
\left[-\frac{1}{2}\left(\frac{p_i-s\,t_i(\delta)}{\sigma_{i}}\right)^2\right],
\end{equation}
where $\sigma$ is the white radiometer noise and both data and
template are baseline-subtracted.  Implicit in this formulation is the
assumption that pulsar emission can be well-described as
amplitude-modulated noise \citep{Rickett75}.  Most pulsars raise the
system temperature negligibly, and $\sigma$ can be taken as a constant
independent of phase and estimated from the baseline variance or
receiver parameters, and this is done in `classical' pulsar timing.
For bright pulsars, especially those we consider here, the peak
intensity may exceed the system equivalent flux density, and $\sigma$
is increased by pulsar `self-noise'
\citep[e.g.][]{Kulkarni89,Gwinn11}, becoming dependent on phase and
unknown \textit{a priori}.  As usual, once an observation provides a
fixed realization of \textbf{p}, $f$ can be viewed as the likelihood
function for the parameter(s), $\mathcal{L}(\delta)$, and an
estimator, $\hat{\delta}$ determined via maximum likelihood.  In
pulsar timing, the likelihood function is typically evaluated in the
Fourier domain \citep{Taylor92}.

In the above formulation, the pdf $f$ is \textit{conditioned} on the
template \textbf{t}, so if our assumptions about \textbf{t} are wrong,
$\hat{\delta}$ will be biased.  In pulsar timing, \textbf{t} is
typically taken to be the mean pulse profile obtained from a long
integration, or a smoothed/analytic approximation thereof.  Because
\textbf{t} changes with each pulse, the observed profile $\textbf{p}$
only approaches the mean asymptotically, with fractional error
$\propto1/\sqrt{t}$.  The increased scatter---jitter/SWIMS---is
a manifestation of the bias incurred by using a time-averaged
template.

There are two fundamentally different paths to dealing with this
uncertainty, both involving
a modification of the likelihood.  As 
presented by \citet{Oslowski11}, one approach is to treat the
pulse-to-pulse variability as correlated noise.  This solution is
elegant as it captures both the increased variance from
phase-dependent system temperature (their `Regime 2') and the
pulse-to-pulse variability (their `Regime' 3).  It comes at the cost
of a more complicated likelihood, as the phase bins are no longer
independent:  
\begin{equation}
\label{eq:correl_like}
f(\mathbf{p}\,|\,\delta,s,\mathbf{t},\mathbf{\sigma}) =
\frac{1}{\sqrt{2\pi\det\mathbf{C}}}\exp
\left[-\frac{1}{2}\mathbf{r^T}\cdot\mathbf{C}\cdot\mathbf{r}\right],
\end{equation}
where \textbf{C} is the non-diagonal covariance matrix of the profile
bins and \textbf{r} are the residuals to the template.  Generally,
\textbf{C} must be measured from the data.  \citet{Oslowski11} 
measured \textbf{C} and used a principle component
analysis (PCA) to recover some of the jitter/SWIMS distortion and
improve the timing precision of PSR~J0437$-$4715 by about 20\%.
\citet{Oslowski13} extended the method to include polarimetry and
achieved a 40\% improvement in precision.

The second approach, which we adopt, views the template itself as an
unknown to be determined from the data.  The latter remain
uncorrelated, though the errors may be heteroscedastic due to
self-noise.  Formally, $t$ is a nuisance parameter in Hilbert space,
and even approximating it as $\textbf{t}$ incurs many additional
degrees of freedom.  Moreover, although the pulse-to-pulse variability
of some pulsars has been studied in detail, it remains a poorly
understood, stochastic process.

In this work, we make the simplifying assumption that a given single
pulse is drawn from a finite family of $N_b$ pulse shapes, $\{b\}$.  For an
integration over many pulses, this is equivalent to expanding $t$ in a
small number of basis functions:
\begin{equation}
\label{eq:basis}
t \approx \sum_{i=0}^{N_b} s_i\,b_i,
\end{equation}
where the $s_i$ will be determined by the number and fluxes of each
pulse family.

The focus of the work below is to test this approach in two scenarios.
In both cases, we have a `training' set of single pulse data from
which a template basis can be constructed.  (We discuss one approach
to basis construction in \S\ref{sec:basis}.)  In the first scenario,
all other timing observations are carried out in `fold mode', in
which many single pulses are coadded.  This configuration is typical
of pulsar timing observations.  In this case, we attempt to
reconstruct the single pulse information by estimating the basis
components of equation \ref{eq:basis} from the fold-mode data.

In the second scenario, we assume we have at least limited access to
single pulse profiles for each timing observation.  (We discuss
precisely what information might be needed in \S\ref{sec:discussion}.)
In this case, we can build up an optimal profile for each observation
by mapping the single pulses to our basis functions, and we find that
such an optimal profile improves the timing precision even further.

To test this method, we use high time resolution observations of Vela,
\vela{}.  Vela is the brightest pulsar in the sky, and like
PSR~J0437$-$4715 is an excellent test case for probing the
jitter/SWIMS limit on current receiver systems.  We describe the
observations and data reduction in \S \ref{sec:observations}.  Both
cases require a template basis, and in \S \ref{sec:basis_construction}
and \$ \ref{sec:flux_tagging} we discuss two effective methods for its
creation.

\section{Observations}
\label{sec:observations}

\begin{figure}
\includegraphics[angle=0,width=0.45\textwidth]{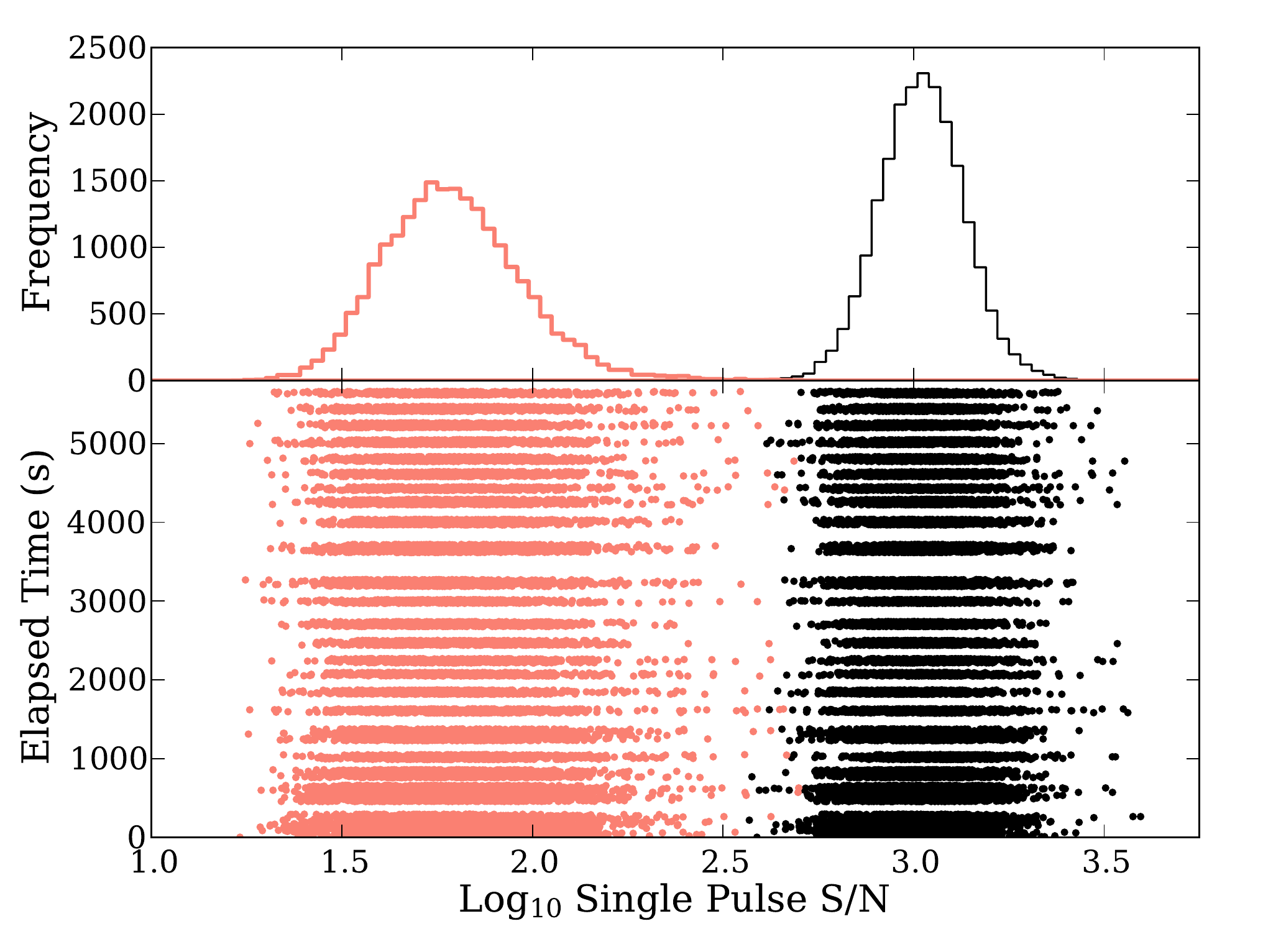}
\caption{\label{fig:sn_histo}The S/N of our single-pulse observations.
The peak S/N for a single phase bin appears to the left in 
salmon, while the total S/N summed over the onpulse region appears to
the right in black.  The bottom panel shows the time elapsed during
the observation and missing intervals; the top panel shows the
cumulative S/N distribution.}
\end{figure}

On 4 July 2012, we used the 64-m Parkes telescope and the CASPSR
backend to record baseband voltage samples from the two orthogonal,
linearly-polarized feeds of a receiver tuned to 2820\,MHz.  The
800\,MHz sampling rate provided useful bandwidth of about 320\,MHz
after filtering. Data were recorded in 8-second blocks, and although
the full observation lasted about two hours, many blocks of data were
lost from the buffer due to the failure of a disk in the on-line
recording system.  Ultimately, we obtained about 30 minutes of data,
or 21,184 pulses, in non-contiguous 8-second blocks (see Figure
\ref{fig:sn_histo}).  We note that 8 seconds is substantially longer
than any reported pulse-to-pulse correlations
\citep[e.g.][]{Krishnamohan83}.

Using \textsc{dspsr} \citep{vanStraten11}, we synthesized coherently
dededispersed filterbanks of the Stokes parameters of each single
pulse with spectral and time resolution of 512 channels and 1024 phase
bins ($\sim$87\,$\umu$s), respectively.  We corrected the filterbanks
for the complex differential gain between the two feeds and for the
bandpass response using observations of a pulsed noise source, and we
finally integrated the filterbanks over frequency and polarization to
produce monochromatic total intensity (stokes I) profiles.  The S/N of
the resulting single-pulse profiles appear in Fig. \ref{fig:sn_histo}.
Vela is truly outrageously bright.

\begin{figure*}
\includegraphics[angle=0,width=0.9\textwidth]{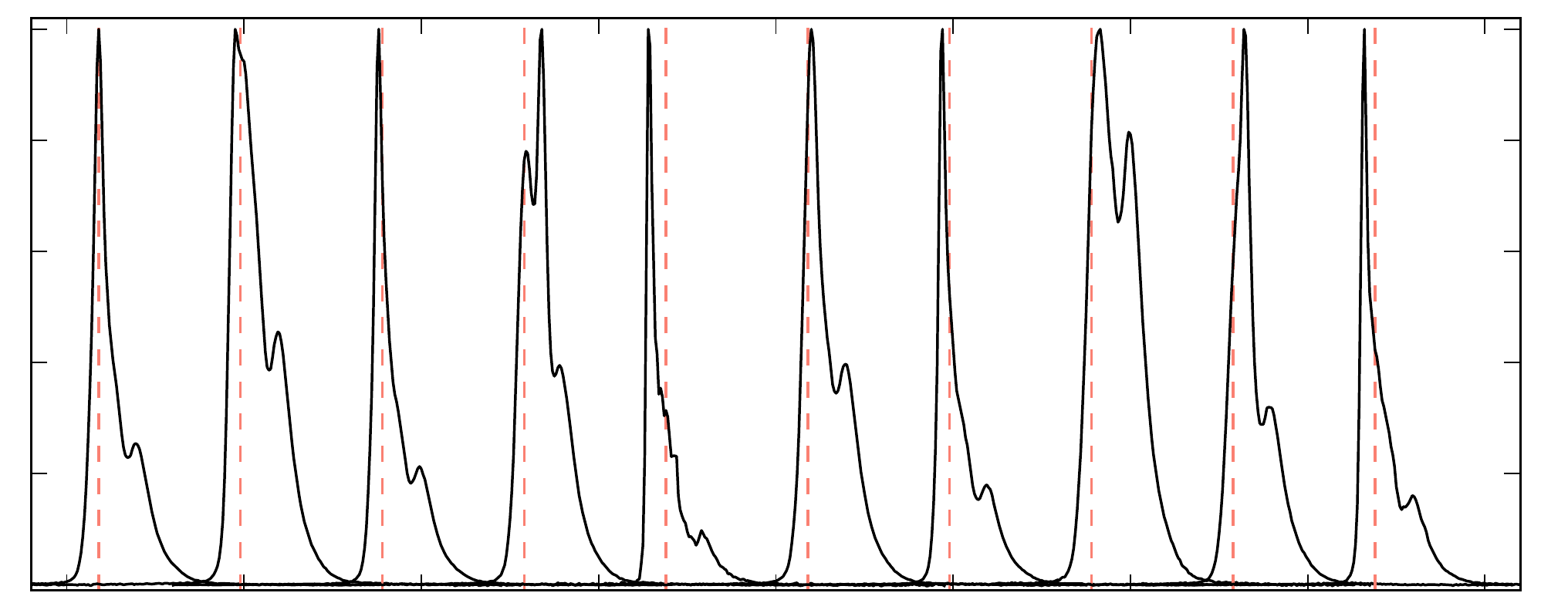}
\caption{\label{fig:basis}An example template basis.  The leftmost
basis function is constructed from pulses most resembling the mean,
while the remainder are based on exceptional single pulses.  The
salmon, vertical dashed lines mark the same fiducial phase, viz. the maximum
intensity in the mean profile.  Several
of the basis functions represent pulses with very bright sub-pulses at
particular phases, while others capture unusual mixtures of
components.}
\end{figure*}

The precise topocentric pulsar frequency at the time of observation is
unknown a priori.  We measured it by extracting a TOA from each pulse
and fitting a linear model.  We computed the phase offset of each TOA
to this linear model and then rotated the pulse profiles by this
offset, equivalent to folding the data at the exact pulsar
frequency.  It is important that this phase alignment matches the true
pulsar rotational phase with adequate precision.  Because we have used
more than $10^4$ pulses to estimate it, the phase alignment of any
individual TOA will be better by a factor of 100 than the TOA
measurement uncertainty.  As we shall see below, this is adequate for our
analysis.

We monitored the gain stability during the observation by analyzing
the baseline (offpulse) noise in our single pulse profiles, dividing
them into 276 (748) onpulse (offpulse) bins.  The mean offpulse level
drifted by about 5\% over the observation, indicating the absolute
gain was stable to at least this level.  Indeed, it was likely more
stable, as we found that the ratio of the mean to the standard
deviation, which is linearly proportional to the system gain,
$\kappa\equiv4.2282(7)\times10^{-3}$, did not vary over the
observation.  We thus expect S/N to be a good proxy for flux density.
We note our observed $\kappa$ is also in agreement band-limited noise,
viz. the radiometer relation, which predicts
$\kappa=4.23\times10^{-3}$ for 320\,MHz of bandwidth.


The dual-band 10/50\,cm \citep{Granet05} observing system we used has
a system equivalent flux density (SEFD) of 25--30\,Jy over the
observing band, and the peak intensity of Vela reaches about 9\,Jy
(mean about 220\,mJy).  Thus, for a typical pulse, the system noise
level is only increased by about 5\% at peak, a negligible effect.
However, peak intensity of the brightest pulses \citep[giant
micropulses, see][]{Johnston01} can be substantially brighter and
exceed the SEFD, increasing the noise by more than 40\%.  We address
this increased noise level further in \S\ref{sec:results}.

Two important systematic effects could affect a single pulse
analysis.  First, broadband RFI can mimic sub-pulse structure.  Such
RFI is easily identified for \vela{} by its lack of dispersion, and
moreover is largely absent from the 10\,cm band; we identified and
discarded only four single pulses affected by impulsive RFI.  Second,
as the antenna tracks the source, the parallactic angle relative to
the feed symmetry axis evolves, changing the illumination of the two
feeds.  Correcting such effects requires sophisticated modeling of the
receiver response \citep[e.g.][]{vanStraten04,vanStraten13}.  But, as
the receiver is co-axial with little cross-polarization and the range
of parallactic angles covered during the observation is small, we
expect this effect to be minimal.

\section{A `Pulse Shape' Template Basis}
\label{sec:basis}

Here, we discuss an approach for constructing a template basis based
primarily on capturing the various observed pulse shapes.  In \S
\ref{sec:flux_tagging}, we discuss an approach based on the flux
density of single pulses.  In general, we expect any such a basis to only
encapsulate a fraction of the true pulse-to-pulse variability.  If
single pulses are actually formed from random realizations of
sub-pulses or `shots' \citep{Cordes75,Hankins03}, there is no
expectation that a family of single pulse shapes should emerge.
Moreover, rare but bright pulses, such as the giant micropulses of
\vela{} or the true giant pulses of PSR~B1937$+$21 \citep{Cognard96},
substantially affect TOA measurements but cannot be faithfully
represented by a small family of basis functions.

Nonetheless, the question is not one of perfection, but of efficacy,
and below we show that a modest basis set can capture a substantial
portion of single pulse variability.

\subsection{Construction}
\label{sec:basis_construction}

An effective template basis is one that reproduces the pulse-to-pulse
variation reasonably well without admitting too many degrees of
freedom.  In principle, for a fixed number of basis functions and a
given set of single pulses, one could determine an optimal basis, i.e.
one that minimizes the mean squared error between each
single-pulse and the best-fitting basis function.  However, this is an
extremely challenging optimization problem!

Instead, we develop a basis iteratively.  The initial basis comprises
the mean template, and we begin by identifying the pulse least similar
to the mean (see below).  We then divide the single pulses into those
more similar to the mean and those more similar to the outlier, and
average those pulses to produce a basis with two templates.  Next, we
select the pulse least like either of these templates, re-classify the
single pulses, produce a basis with three templates, and so on and
so forth.

Generally, single pulse flux densities follow a lognormal
distribution \citep{Burke-Spolaor12}, while some pulsars displaying giant pulses and
micropulses follow power law statistics, \citep[see,
e.g.,][]{Cairns01,Cairns04}.  To avoid constructing our
basis from rare bright pulses, which would naturally be the strongest
outliers due to their high S/N, we adopt a `reduced' $\chi^2$,
\begin{equation}
\chi^2_r\equiv\sum_{i=1}^{N_p}(p_i-t_{i})^2/\sum_{i=1}^{N_p}p_i^2,
\end{equation}
which measures the distortion between the observed pulse shape and the
template normalized to the S/N.

We note that this procedure does not necessarily converge to a unique
basis.  With each iteration, the basis functions are revised, and
profiles `belonging' to one basis function may be shuffled to
others during the next classification.  In practice, we find the the
functions and classifications settle to a stable result after a few
iterations.

To construct a mean pulse shape template, we co-added the first $10^4$
Vela pulses.  We then applied the basis construction algorithm
outlined here to the same pulses to compute template bases with
varying numbers of basis functions.  Templates from an example
resulting basis appear in Figure \ref{fig:basis}.

\subsection{Application and Efficacy}
\label{sec:basis_sims}

\begin{figure}
\centering
\includegraphics[angle=0,width=0.45\textwidth]{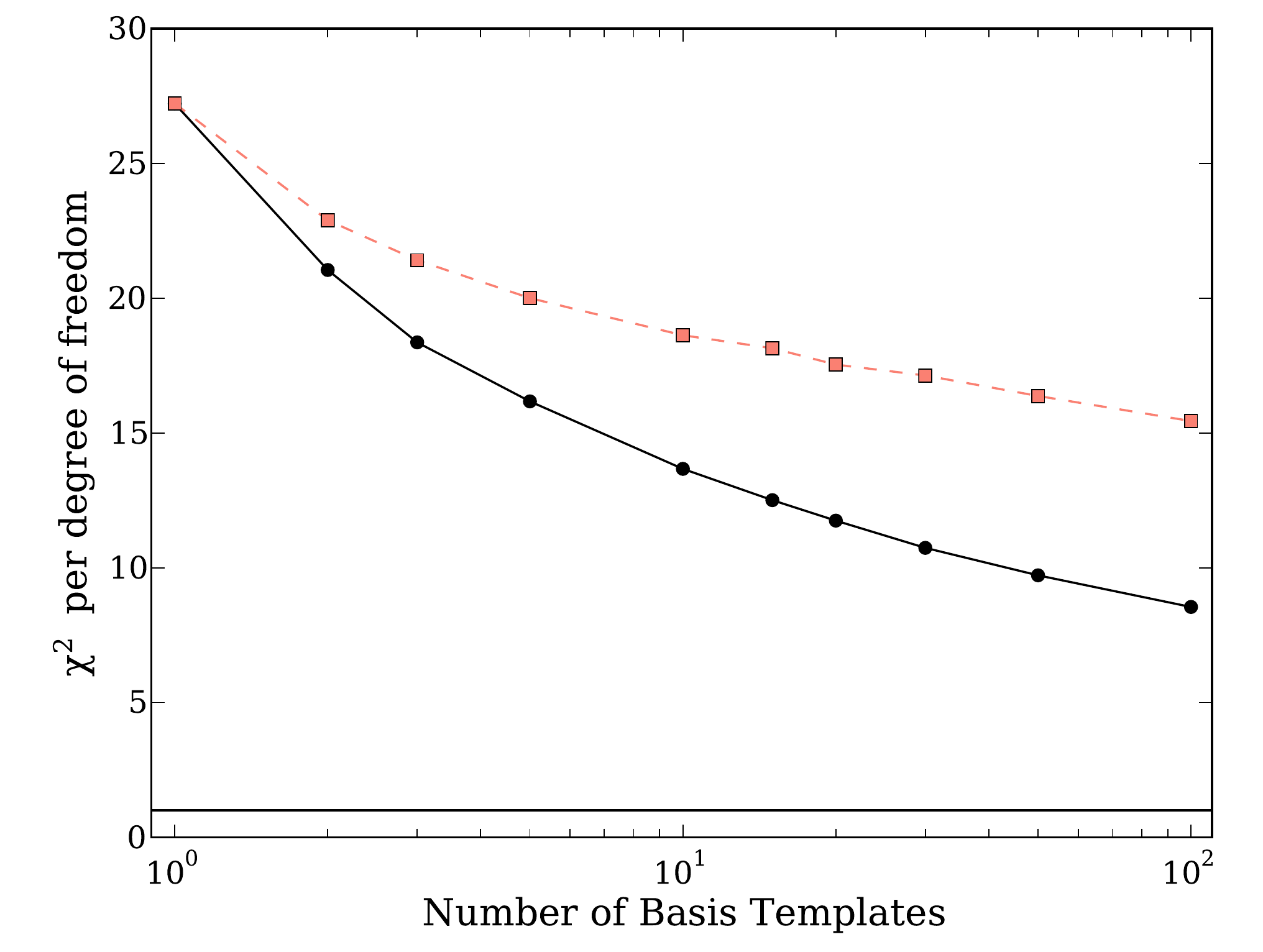}
\caption{\label{fig:chi2_red_vs_nb}The total mean squared error
($\chi^2$), normalized by the total degrees of freedom, for the $10^4$
single pulses as a function of the number of functions in the
template basis.  The black solid line traces shows the `pulse shape'
basis (\S \ref{sec:basis}) and the $\chi^2$ scales approximately as $N_t^{-1/4}$.  The
salmon, dashed line shows the same trend for the `flux density' basis
of \S \ref{sec:flux_tagging}.}
\end{figure}

Here, we explore how well the basis encapsulates the pulse-to-pulse
variability.  First, we analyze the reduction in total mean squared
error
($\chi^2$) of the set of single pulses relative to the mean
template.  For each pulse, we compute $\chi^2$ relative to either
the mean template or the best-fitting basis template, and we then
total the $\chi^2$ over all pulses.  The results, shown in Fig.
\ref{fig:chi2_red_vs_nb}, indicate a dramatic improvement in $\chi^2$ with
only a few basis functions.  In fact, the total $\chi^2$ scales
roughly as $N_t^{-1/4}$, though this relation is certainly
coincidental and must steepen before $N_t$ becomes equal to the total
number of single pulses!

\begin{figure}
\centering
\includegraphics[angle=0,width=0.45\textwidth]{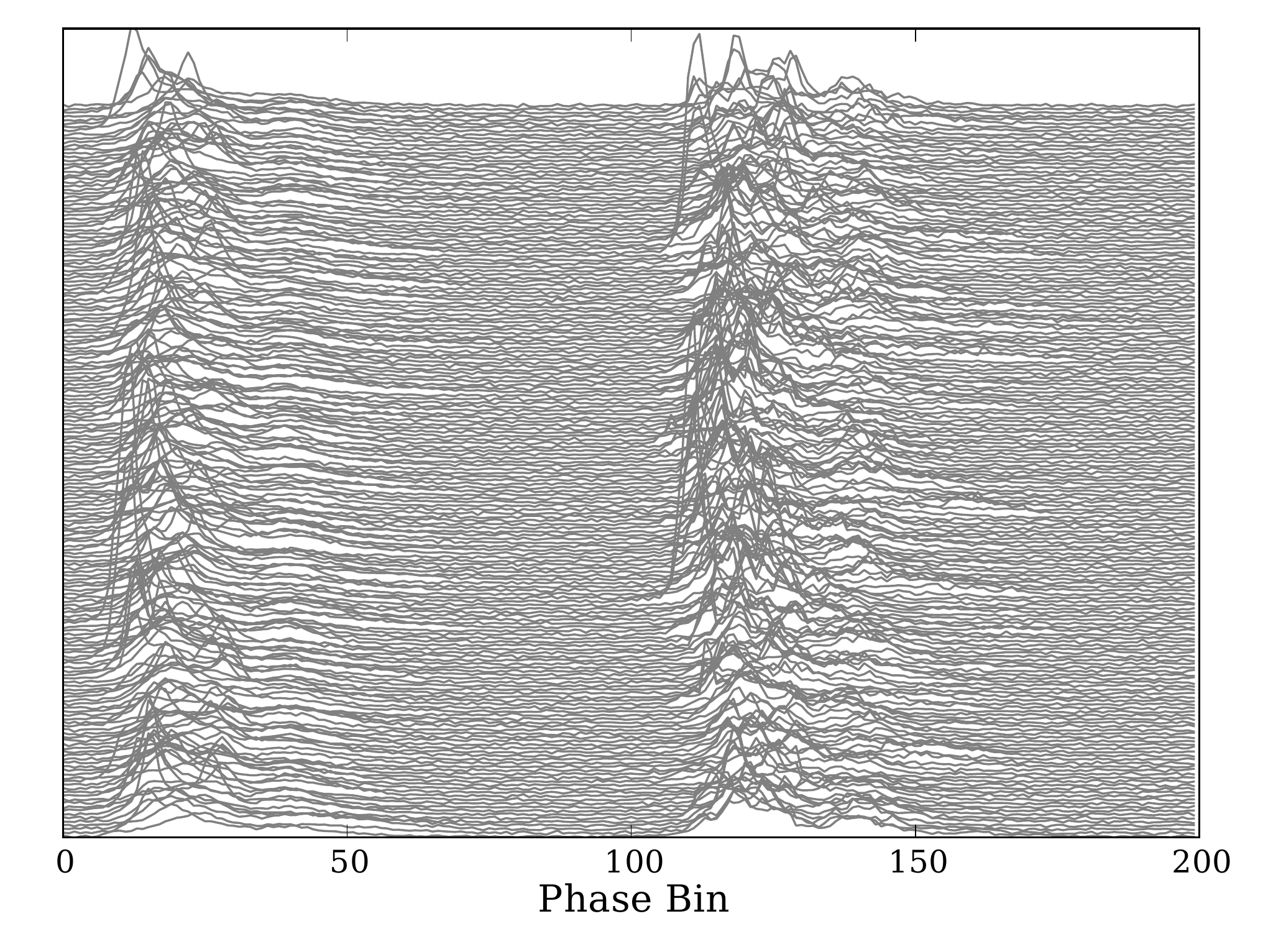}
\caption{\label{fig:pulse_train}An example of 200 single pulses
simulated as described in \S\ref{sec:basis_sims}.  Simulated pulses
appear on the left, while 200 consecutive single pulses from
\vela{} appear to the right.}
\end{figure}

It is implicit in our definition of a single-pulse template basis that
each pulse corresponds to one and only one basis function.  We further
make the assumption that pulses are uncorrelated.  This is not true on
short time-scales \citep[a few pulses,][]{Krishnamohan83}, but is a
good assumption for longer intervals.  (We discuss pulse-to-pulse
correlation and its mitigation further in \S\ref{sec:discussion}.)
With these assumptions, it is straightforward to simulate pulse trains
and compare these with observations.  First, we classify the $10^4$
single pulses according to basis function by minimizing $\chi^2$.  The
fractions of pulses falling in each class are just the parameters of a
multinomial distribution.  Second, we have observed in our sample that
that the distribution of pulse flux densities within each class is
approximately described by a lognormal distribution.  Thus, we realize $N$
simulated pulses by drawing an $N$-pulse sample from a multinomial
distribution and then, for each classified pulse, drawing a flux from
the appropriate lognormal distribution.  An example train of 200
single pulses is shown alongside those taken from real data in Fig.
\ref{fig:pulse_train}.

\begin{figure}
\centering
\includegraphics[angle=0,width=0.45\textwidth]{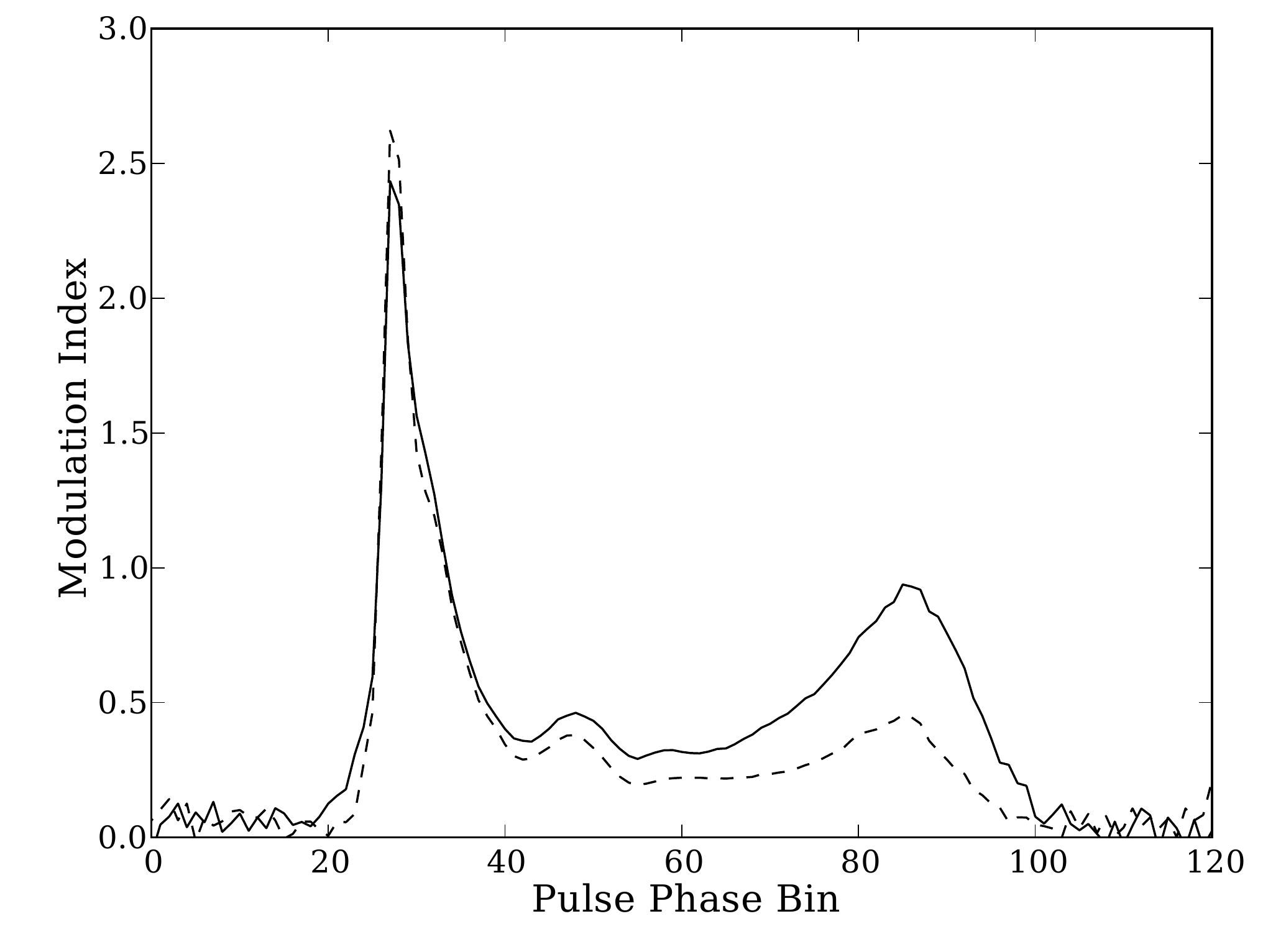}
\caption{\label{fig:prof_sim1}The phase-resolved modulation index (see
main text) of the flux density of $10^4$ single pulses from
simulations (dashed) and data (solid).}
\end{figure}

\begin{figure}
\centering
\includegraphics[angle=0,width=0.45\textwidth]{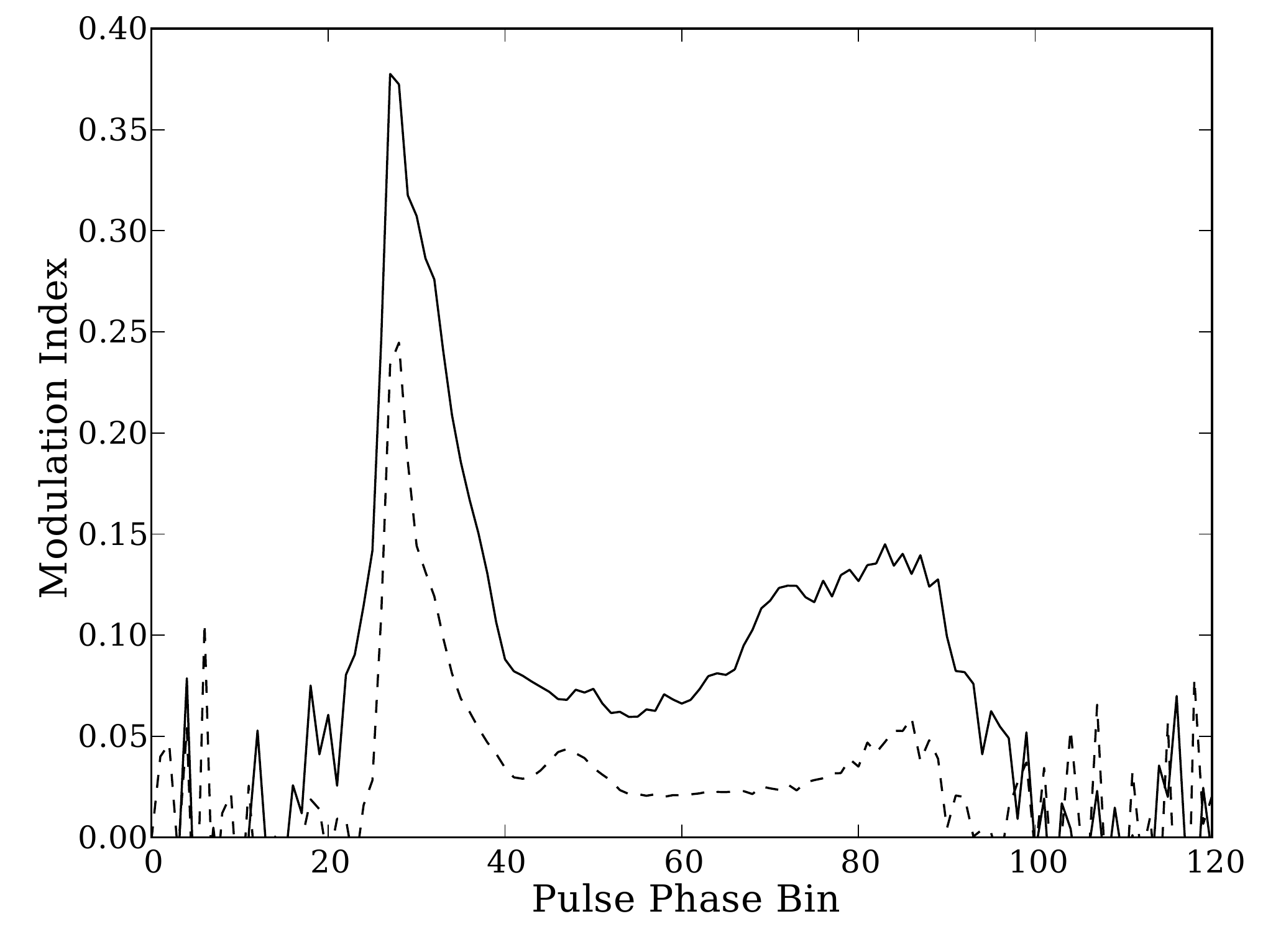}
\caption{\label{fig:prof_sim100}The phase-resolved modulation index
of the flux density of 100 100-pulse sub-integrations from simulations
(dashed) and data (solid).}
\end{figure}

Qualitatively, the simulations do a reasonable job of reproducing the
observed pulse trains, though the real pulses show more stucture,
particularly along the trailing edge of the peak.  To quantify this,
we have computed the modulation index \citep[see][for
discussion]{Jenet01}, defined as
$m=\sqrt{\sigma^2(\phi)-\sigma^2_{\mathrm{off}}}/\mu(\phi)$, i.e. the
excess variance scaled to the mean intensity as a function of pulse
phase.  Figure \ref{fig:prof_sim1} shows the results for single
pulses: most of the modulation in the leading peak is captured, while
variations are in the trailing region of the pulse in the real data
are roughly twice those of the simulations.  Much more interestingly,
Figure \ref{fig:prof_sim100} shows a similar factor of two deficit in
simulation variability for 100-pulse sub-integrations.  While some of
this discrepancy results from failing to include self-noise in the
simulations, the dominant effect must be pulse-to-pulse correlation,
e.g. as observed by \citet{Krishnamohan83}.  On the other hand, we
expect the simulations to become more faithful (approaching the
single-pulse results) for longer sub-integrations.

\section{Results}
\label{sec:results}

Here, we describe the application of three different timing algorithms
and the `pulse shape' basis.  The first algorithm, `profiling', does
not require single pulse data, and we use it to introduce our analysis
framework.

\subsection{Applications without single pulse data -- `P'}
\label{sec:prof}

In the most common mode of pulsar observation, the pulsar signal time
series is folded at the topocentric rotational frequency, producing a
sub-integration containing many co-added single pulses.  A typical
integration time of 30\,s records of order 100 pulses from a garden
variety unrecycled pulsar.  While the information contained in the
single pulses is lost, we can still make inferences about the single
pulse content of the sub-integration by estimating the coefficients
of the basis template.  Those inferences in turn can furnish improved
constraints on the phase offset $\delta$.  In terms of the basis
function coefficients, the pdf for the sub-integration
profile $\textbf{p}$ becomes
\begin{equation}
\label{eq:prof_like}
f(\mathbf{p}\,|\,\delta,\mathbf{s},{\mathbf{b}},\mathbf{\sigma}) =
\prod_{i=1}^{N_b}\frac{1}{\sqrt{2\pi}\sigma_i}\exp
\left[-\frac{1}{2}\left(\frac{p_i-(\mathbf{s}\cdot\mathbf{b})_i(\delta)}{\sigma_i}\right)^2\right].
\end{equation}
For pulsar timing, the basis coefficients \textbf{s} are nuisance
parameters.  To maintain similarity with current pulsar timing
techniques, we choose to use the \textit{profile likelihood} to
estimate $\delta$.  That is, for each trial value of $\delta$, we determine
the value of \textbf{s} that maximizes the likelihood, reducing the
likelihood to a one-dimensional function $\mathcal{L}_p(\delta)$.
Because \textbf{s} is formally positive semidefinite (no negative flux
densities!), we restrict the optimization of \textbf{s} to positive
values.

In the presentation below, we ignore the pulsar self-noise
contribution and assume a constant $\sigma$.  We repeated our analysis
including self-noise in the model, but found negligible changes to the
measurements of $\delta$ and its uncertainty.  That is, for
Vela, the self-noise effect is small compared to pulse-to-pulse
variability.  As introducing the self-noise (heteroscedasticity)
substantially complicates the computation for little gain, we neglect
it henceforth.

\begin{figure}
\centering
\includegraphics[angle=0,width=0.45\textwidth]{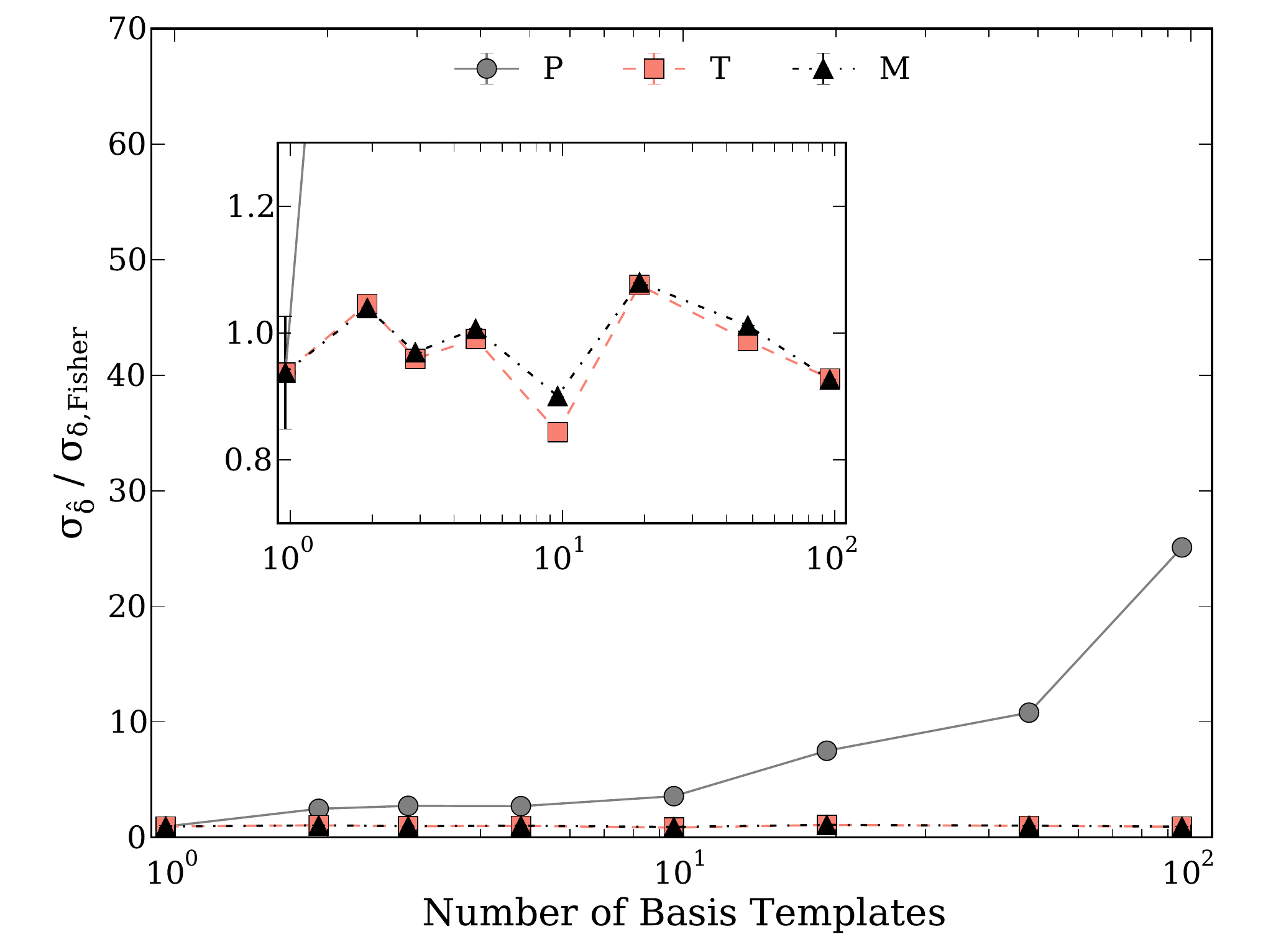}
\caption{\label{fig:mse_vs_nb_sim1}The precision of the phase
measurement $\delta$ as a function of the complexity of the basis for
three different realizations of the likelihood, using 100-pulse
sub-integrations  of \emph{simulated} data.  The grey, solid trace
(`P') corresponds to the profile method, discussed in
\S\ref{sec:prof}.  The dashed salmon and dot-dashed black lines
correspond to the `tagging' methods discussed in \S\ref{sec:tag}, both
without (`T') and with (`M') the addition of the multinomial
likelihood.  All results are scaled to the ideal measurement precision
(see main text).  The inset is as the main figure, but zoomed to show
that the tagging methods perform optimally, reaching the Fisher
precision.  In these results, the abscissa indicates $N_b$ for both
the simulations and the fitting, i.e. as if one had perfect knowledge
of the family of single pulse variations.  Finally, note that the
error bars due to finite sample size are shown for the $N_b=1$ and
suppressed for the remainder; see main text for details.}
\end{figure}

\begin{figure}
\centering
\includegraphics[angle=0,width=0.45\textwidth]{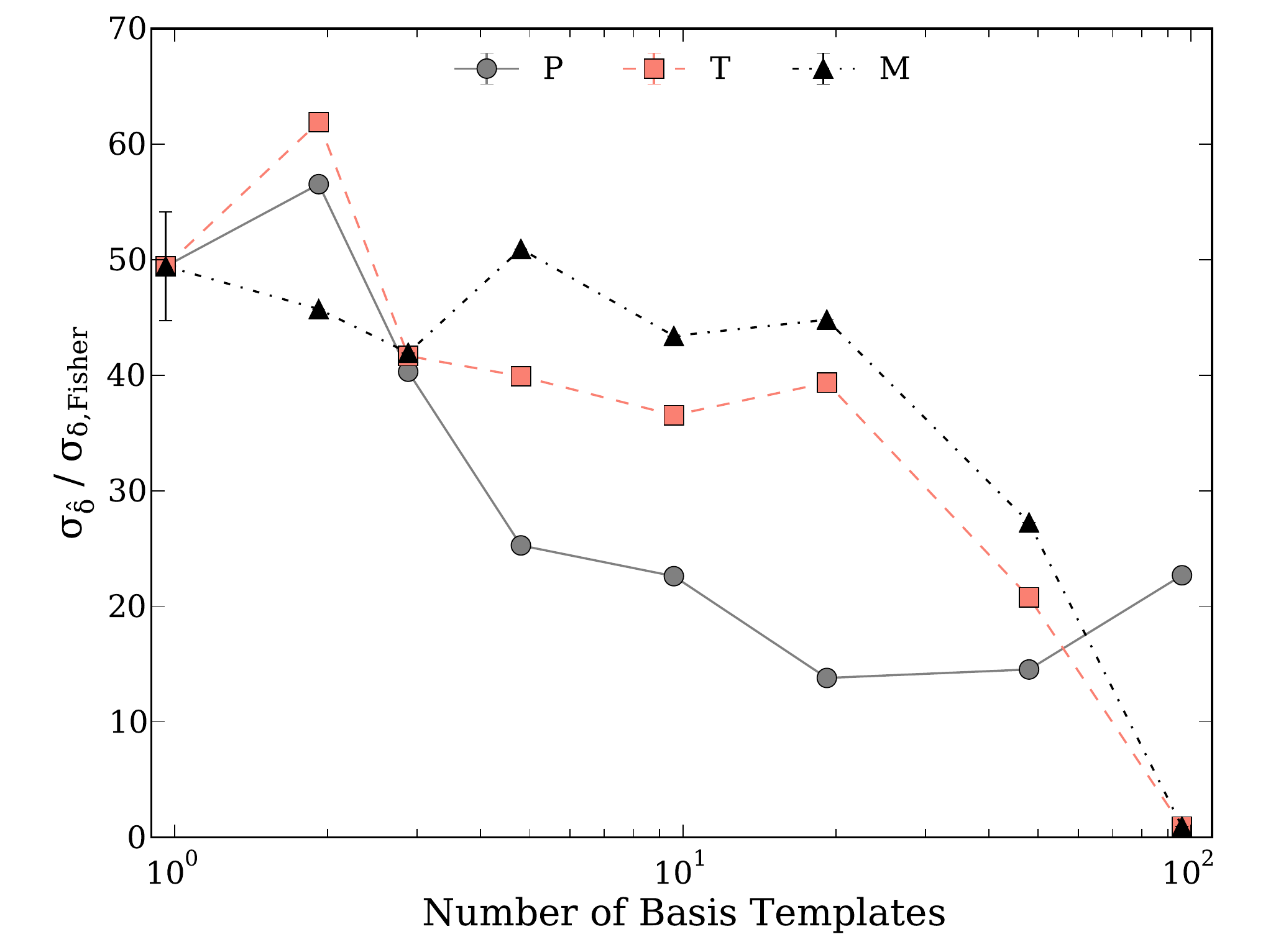}
\caption{\label{fig:mse_vs_nb_sim2}As Fig. \ref{fig:mse_vs_nb_sim1},
but the \emph{simulations} use an $N_b=100$ basis while the abscissa
indicates the $N_b$ used in the fits.  This case corresponds to
imperfect knowledge of the panoply of pulse shapes.}
\end{figure}

\subsubsection{Simulations}
To validate and characterize the approach under controlled but
realistic conditions, we simulated pulse trains from the basis as
described in \S\ref{sec:basis_sims} and then synthesized
sub-integrations with 100, 200, and 400 pulses.  These lengths, and
the number of simulated sub-integrations, are chosen to match those
of the data, described below.

For each sub-integration, we maximized the likelihood of equation
\ref{eq:prof_like} to determine the phase shift estimator
$\hat{\delta}$.  We then computed the standard deviation of
$\hat{\delta}$ over the set of simulations and normalized it to the
idealized scatter, i.e. what we would observe in the absence of
jitter/SWIMS and self-noise.  This uncertainty is given by the Fisher
information of the mean template, which gives a lower limit for
the variance for $\hat{\delta}$,
\begin{equation}
\label{eq:fisher}
\sigma^{-2}_{\hat{\delta},\mathrm{Fisher}} = \sum_i\left(\frac{1}{\sigma_r}\frac{\partial t_i}{
\partial \delta}\right)^2,
\end{equation}
where $\sigma_r$ is again the radiometer noise.

In our simulations, we can choose to simulate profiles with a basis
with the same $N_b$ we use to maximize the likelihood, or we can fix
it at some representative $N_b$.  The former essentially validates the
method, while the latter probes how the results change when the basis
and the data disagree, as will certainly be the case for real data.
Accordingly, we follow both approaches.

The results for which $N_b$ is the same for both simulations and
fitting appear in the grey traces labelled `P' in Fig.
\ref{fig:mse_vs_nb_sim1}.  Here, for clarity, we show the results for
the 100-pulse sub-integrations; the 200-pulse and 400-pulse results
are similar. The leftmost values, corresponding to a template basis
with $N_b=1$, i.e. the mean template or `traditional' pulsar timing,
shows a scatter consistent with the Fisher limit---certainly to be
expected, since simulations from a template with $N_b=1$ have no
jitter/SWIMS.  Since the set of simulations changes for each $N_b$,
the data are uncorrelated; this will not be the case for scenarios
described below.  As the complexity of the simulations increases, we
see the profile method slowly diverges as the degrees of freedom
increase.  Thus, we expect the method may ultimately break down if we
attempt to fit a pulsar with very complicated pulse-to-pulse
variability.

In the second set of simulations, we fix $N_b=100$ for a high level of
pulse-to-pulse variability, and we then vary $N_b$ in our fits of
$\hat{\delta}$.  Now, the $N_b=1$ case shows a jitter/SWIMS level of
about 50 times the radiometer limit, consistent with our observation
that our simulations produce about half of the pulse-to-pulse
variability of the actual data.  As we increase $N_b$, the
profile method is able to successfully reconstruct some of the single
pulse variability, reducing the jitter/SWIMS level by a factor of
three for $N_b=20$, before slowly diverging.  Since we use the same
set of simulations for each value of $N_b$ on the abscissa, these
points are correlated, and the error bars should be interpreted as
uncertainty on overall amplitude rather than as a measure of scatter.

Thus, we conclude that this `profile' method is quite effective in
removing jitter/SWIMS noise under the idealized conditions of our
assumptions about single pulse variability.  This approach should be
especially effective for pulsars with less complicated pulse-to-pulse
variability, and for those whose single pulses are too faint for
effective single pulse `tagging' (see below), such as MSPs.

\begin{figure}
\centering
\includegraphics[angle=0,width=0.45\textwidth]{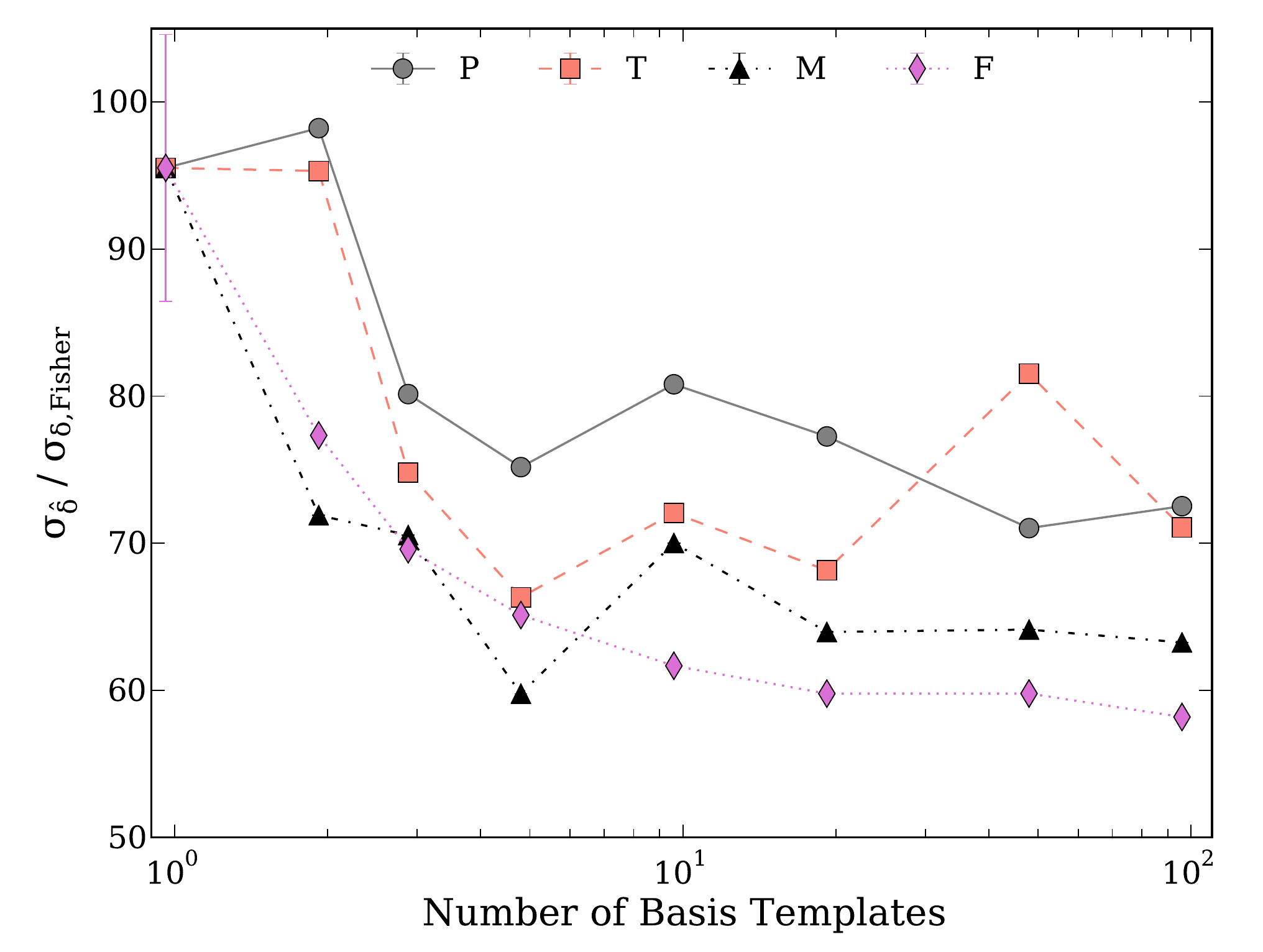}
\caption{\label{fig:mse_vs_nb_data}As Fig. \ref{fig:mse_vs_nb_sim1},
but for \emph{data} from \vela{}.  Additionally, the orchid, dotted trace
shows the results for the flux-tagging scheme described in
\S\ref{sec:flux_tagging}.}
\end{figure}

\subsubsection{Data}

We now apply the profile method to synthesized sub-integrations of
varying lengths from the data set described above.  After excising the
$10^4$ pulses used to build the basis template and those affected by
RFI, 11,182 pulses remain.  Maintaining a sufficiently large sample of
sub-integrations to measure the scatter of $\hat{\delta}$ imposes a
limit on sub-integration length of about 400 pulses, or 36\,s,
typical of a fold-mode sub-integration.  However, as we discuss below,
we expect the results to scale to longer integrations, and perhaps to
improve as correlations between single pulses become less important.

As with the simulations, for each synthesized sub-integration, we
maximized $\mathcal{L}_p(\delta)$ to determine $\hat{\delta}$ and
examined the distribution of values.  The results are summarized in
Fig. \ref{fig:mse_vs_nb_data}.  As before, the leftmost points
correspond to `traditional' pulsar timing and show the jitter/SWIMS
level is about 100 times the ideal limit.  As with the second set of
simulations, because the same data
set is used for each set of basis templates, the results are
correlated.

As we might expect from the failure of our basis to reproduce the
lion's share of the pulse-to-pulse variability, the reduction in
scatter in moving to $N_b>1$ is substantially poorer than in
simulations.  Nonetheless, we see a gradual impovement in the level of
scatter, as the basis complexity increases and captures more of the
pulse-to-pulse variability, with a typical scatter reduction of about
25\%.  We also note that longer sub-integrations show a
lower level of jitter/SWIMS relative to the ideal (which scales as
$1/\sqrt{t}$) than shorter sub-integrations, suggesting that
performance does indeed improve as the pulse-to-pulse correlation
time-scale becomes short compared to the sub-integration length.  

\begin{figure}
\centering
\includegraphics[angle=0,width=0.45\textwidth]{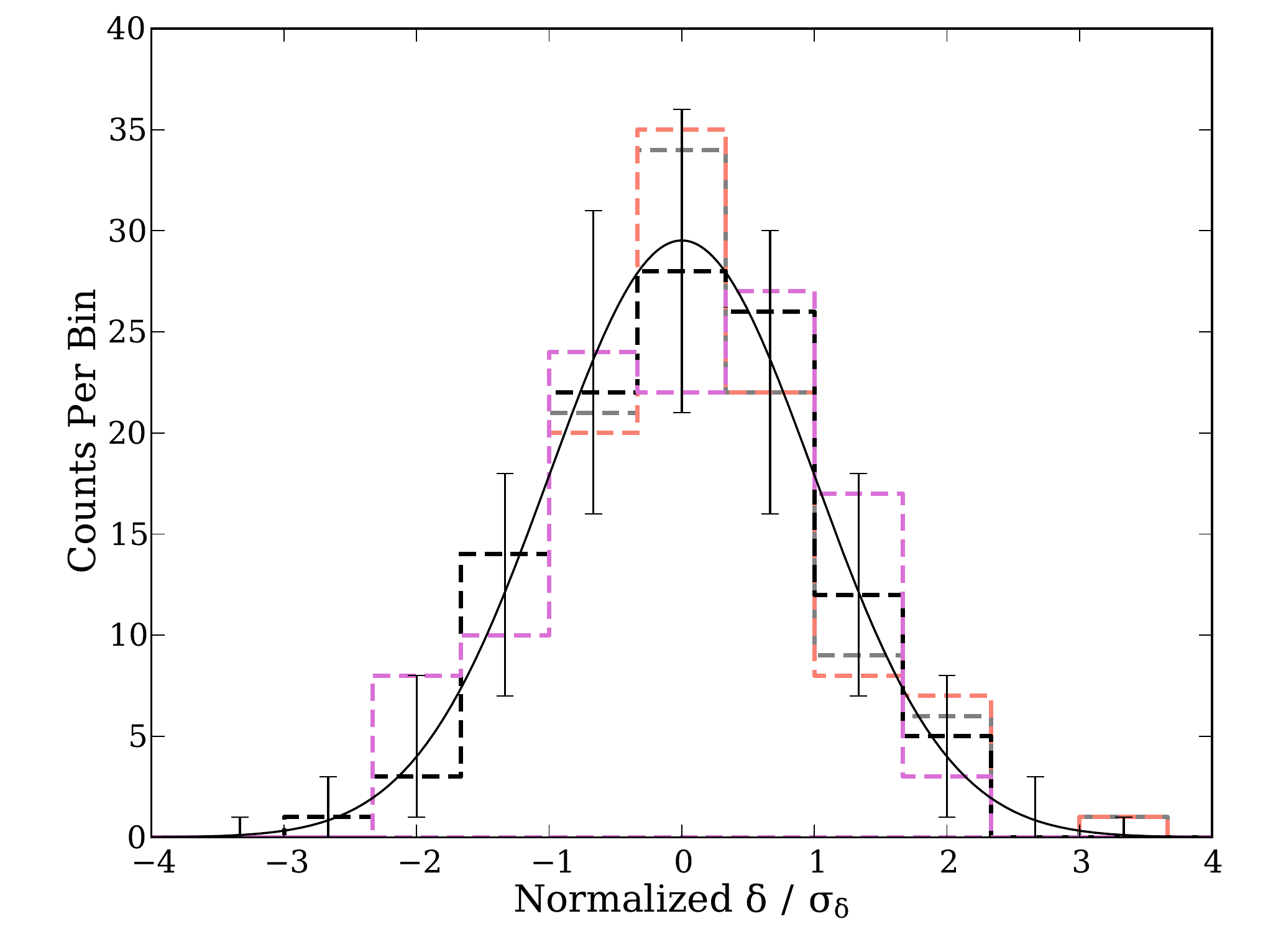}
\caption{\label{fig:pull_hist}The distribution of the error-normalized
residuals after normalization by the sample standard deviation, for
all three `shape' algorithms, with 100-pulse sub-integrations and
$N_b=20$.  The black trace and error bars show the
expected gaussian distribution and the range covered by 90\% of random
realizations of identical histograms of gaussian random variables.
The colors follow the convention of Fig.
\ref{fig:mse_vs_nb_sim1}.  The orchid line shows the same result for
the `flux basis' of \S \ref{sec:flux_tagging}.}
\end{figure}

While a reduction in the scatter of measured phase offsets clearly
improves precision, an accurate estimate of the remaining noise is of
vital importance for pulsar timing.  Ideally, reliable error estimates
can be extracted directly from the timing procedure, and we expect
this could be achieved even without a perfect template basis via
bootstrap methods.  But minimally, all we require is that the
normalized residuals follow a gaussian distribution.  These normalized
residuals appear in Fig. \ref{fig:pull_hist}, showing that the
estimators for $\hat{\delta}$ do indeed follow a normal distribution.

\subsection{Applications with single pulse data -- `T' and `M'}
\label{sec:tag}

Next, we consider a hybrid mode of observation---one in which we
produce sub-integrations as before, but in which we also have access
to single pulse data.  While many avenues are available, including
marginalization over template classification and Monte Carlo Markov
chain methods, we opt for a simple approach that might more readily be
implemented in an online pulsar data recorder.  Because nearly every
pulse in our data set has a high S/N, it can be uniquely associated
with a single basis function.  This leads to the idea of `pulse
tagging'---each incoming single pulse is classified and its flux
(scale parameter) measured; the pulse itself can then be discarded.
The classifications and scales are then used to build an optimized
template from the template basis, which is in turn used to measure the
phase offset of the full sub-integration.  A key point: we are not
``using the data to fit the data''.   Because we do not know the correct
phase offset $\delta$ \textit{a priori}, we perform classification and
template construction over a range of values of $\delta$.  This
process is similar to constructing a profile likelihood, in which we
optimize for the (discrete) profile classification parameters.

\subsubsection{Simulations}

As before, we test this method with simulated data.  In the first
`sanity check' approach, we see that the approach performs
flawlessly.  That is, by tagging the individual pulses, we are able to
construct a template for each sub-integration that entirely accounts
for jitter/SWIMS, and we reach the Fisher limit for estimator
uncertainty.  Importantly, despite implicitly fitting for the profile
class and scale of each pulse, we see no evidence for overfitting.

The second set of simulations, in which all simulated profiles are
drawn from a $N_b=100$ template basis, are more interesting.  For
fits with $N_b\ll100$, there is no improvement in jitter/SWIMS
level.  But once we reach $N_b=50$, jitter/SWIMS is reduced by roughly
half, and as in the first set of simulations we remove jitter/SWIMS
entirely once $N_b$ is consistent between simulation and fit.  This
result is readily explained.  For the $N_b\ll100$ cases, no template
basis function maps well onto a single pulse, and tagging process will
essentially build up the mean template.  On the other hand, once the
basis begins to encapsulate single pulses, we achieve a template
that accurately records the influence of pulse-to-pulse variability.

From these results, we conclude that the single pulse tagging approach
is extremely effective (entirely removing jitter/SWIMS) if the
template basis is well-suited to the data.

\subsubsection{Data}

The results of applying this approach to our test data set appear in
the salmon traces of Fig. \ref{fig:mse_vs_nb_data}.  Jitter/SWIMS
decreases rapidly with increasing $N_b$ through $N_b=5$, before
saturating at roughly 65\% of the $N_b=1$ level.  These results are
largely consistent with our understanding that the template basis
should be well matched to single pulse variability.  As we saw in Fig.
\ref{fig:chi2_red_vs_nb}, we capture most of the pulse-to-pulse
variability \emph{that we can} with relatively few template basis
functions, and increasing $N_b$ beyond about 5 yields only slow gain. 

Although the tagging method seems effective in simulations, there is
in principle additional information we can apply to the fits that may
better handle real data.  E.g., very rare, bright pulses can
substantially affect a TOA measurement \citep{Oslowski14}, so they
must be represented in an effective template basis.  On the other,
less intense pulses with morphology similar to giant pulses should not
be so classified.  Put another way, the pulse counts and flux
densities associated with each basis are random variables, and we can
add terms reflecting this to the likelihood.  Substantial
misclassification will then be penalized.  As discussed in
\S\ref{sec:basis_construction}, we can easily determine the expected
rates of pulses associated with each basis, as well as measure their
flux distribution.  With our neglect of pulse-to-pulse correlation,
these terms enter the likelihood multiplicatively through the
multinomial distribution (for the classes) and $N_p$ lognormal
distributions (for the fluxes).  Suppressing constant terms, this log
likelihood is
\begin{equation}
\label{eq:like_mn}
\begin{split}
\log\mathcal{L} = &-\frac{1}{2}\sum_{i=1}^{N_p}
\left[p_i-t_i(\delta)\right]^2 + \left[\frac{\log
s_i-\mu_b}{\sigma_b}\right]^2 +2\log s_i 
\\ &+ \sum_{i=1}^{N_b} \left[ n_i \log a_i -\log n_i!\right],
\end{split}
\end{equation}
where
\begin{equation}
\mathbf{t}=\sum_{i=1}^{N_p}s_i\mathbf{b}_i,
\end{equation}
$\mathbf{b}_i$ is
the best-fit basis for the $i$th pulse, and $s_i$ is the corresponding
best-fit flux.  The $a_i$ are the parameters of the multinomial
distribution (relative pulse frequency for each basis) and the $n_i$
are the number of pulses tagged into each basis.

In our data---and in any data in which jitter/SWIMS is
important---the `$\chi^2$' term in equation \ref{eq:like_mn} utterly
dominates the multinomial and lognormal terms, and the relative
weighting must be adjusted to allow the latter data to provide
constraints.  There are two `natural' approaches, and we find they
yield similar results: (1) re-scale the $\chi^2$ term such that the
residual $\chi^2$ per degree of freedom is unity or (2) re-scale the
$\chi^2$ such that the $\chi^2$ term contributes equally to the total
likelihood.  We adopt the latter approach.

Applying these additional constraints yields a small
decrease in the scatter of our sample, in some cases reducing the
jitter noise by as much as 40\%; see the black traces of Fig.
\ref{fig:mse_vs_nb_data}.

\subsection{A Flux Basis}
\label{sec:flux_tagging}

The previous methods have all focussed on pulse \emph{shape} rather
than information encoded in the pulse profile normalization, i.e. its
flux density.  However, for \vela{}, there is a additionally a strong
correlation between the peak intensity of a pulse and the phase at
which that peak falls \citep{Krishnamohan83}, and strong pulses have a
large influence on TOA estimation.  We thus consider here a `flux
basis' and investigate its performance relative to the `shape
basis'.

We construct this basis by simply computing the peak S/N in each
single pulse, dividing the pulses into bins logarithmically uniform in
peak S/N, and constructing templates from the average pulse shape in
each bin.  To perform timing on a synthesized sub-integration, we
assume as in \S\ref{sec:tag} that we can measure single pulse
properties, specifically the peak S/N in each single pulse, selecting
the basis template from the matching S/N bin, and adding up the
matched templates.  The resulting template naturally reflects the
outsized contribution of strong fluxes to a sub-integration.  Note
that it also establishes the additional observational requirement of
stable gain/flux calibration.

The results are extremely interesting.  In Figure
\ref{fig:chi2_red_vs_nb}, we note that the flux basis captures
less of the pulse-to-pulse variability.  The timing performance,
however, is excellent.  The results appear as the orchid-coloured
trace labelled `F' in Figure \ref{fig:mse_vs_nb_data}, showing a
performance that is generally the best of the methods explored here,
offering a robust 40\% improvement over the jitter/SWIMS noise level!

The root of this discrepancy between superiour timing performance and
poor pulse-shape reproduction offers some insight into the Vela's
pulse emission process.  Aside from the clear flux/phase correlation,
it is evident that the emission mechanism can produce pulses with
similar shapes at different observed phases, which in turn may point
to propagation effects as an important source of jitter/SWIMS.
Whether similar effects are common in other pulsars, and how they
might be explored (e.g. through polarimetry) are interesting topics
for future work and we discuss them further below.

\section{Summary and Future Work}
\label{sec:discussion}

We have presented a new method for pulsar timing using single-pulse
based likelihood.  In simulations, we showed these methods are capable
of materially reducing or entirely eliminating the jitter/SWIMS noise
associated with pulse-to-pulse variability.  We demonstrated less
dramatic but nonetheless substantial reductions in the jitter/SWIM
noise of data from the bright Vela pulsar.

Our results are encouraging and motivate future work.  In particular,
much could be done simply be expanding the scope of the data used to
build the basis function.  Our `training set' of pulses for Vela was
limited to a mere $10^4$ single pulses, comparable in size to the set
of pulses we could use to test the algorithms.  It must necessarily
have missed rare pulses, and the limited number of basis function
likely forces the combination of similar but distinct pulse classes
into a single template.  Indeed, our simulations, in which we had
perfect control over the complexity of pulse-to-pulse variability,
showed that we required a basis with nearly the full complexity of the
simulations to fully remove jitter/SWIMS (Figure
\ref{fig:mse_vs_nb_sim1}).

Moreover, Vela exhibits pulse-to-pulse correlation, contrary our
simplifying assumption of independent pulses.  Though more complicated
computationally, a `two-pulse' template basis---whose members are two
consecutive single pulses---is straightforward conceptually and could
capture much of the observed single pulse correlation.  Extension to
`N-pulse' bases would help even more. Both cases would require more
extensive training sets.  Likewise, it is straightforward to extend
these methods to pulsars with periodic sub-pulse modulation, in which
case the natural unit of a basis is the modulation period.

In general, it seems likely that per-pulsar tuning will be required to
find the best combination of template basis and fitting method.  For
Vela, an optimal method must clearly take advantage of the TOA / flux
density correlation.  In the case of PSR~J0437$-$4715,
\citet{Oslowski13} observed a correlation between the phase centroid
and the pulse polarization, suggesting an optimal template basis for
thise pulsar must include polarimetry.  For mode changing pulsars, a
separate template basis for each mode could be determined, and
subsequent application of the timing algorithms would furnish both
correct TOAs and an automatic classification of the pulsar modes.

While there is much future work to be done, a pressing question
remains: if the current template basis can do a good job of describing
single-pulse variation (Figure \ref{fig:chi2_red_vs_nb}), why is the
timing only modestly improved?  This ultimately boils down to the
still poorly understood phenomenology of single pulse variability.
For pulsars whose emission truly comprises `shots' of emission within
a guiding envlope \citep{Cordes75}, fully modelling the covariance
\`{a} la \citet{Oslowski11} is the optimal approach.  Whereas pulsars
which demonstrate at least some underlying pattern, in which similar
single pulses appear repeatedly over time, will benefit from the
recovery of additional information by the deterministic approach
advocated here.  Though the pulse-to-pulse correlation observed for
\vela{} implies at least some repetition, it may yet be that the
single pulses also possess an irreducible random component.  In this
case, the reduction of jitter/SWIMS by 30--40\% achieved here may be
the best possible, and the rest that can be done is accurately
estimate uncertainties.  Indeed, improving the timing algorithms in
the ways outlined above will provide a stringent test of the true
randomness of the pulsar mechanism, and it is thus of keen interest to
collect single pulses from a large set of bright pulsars for further
study.

What would be required to implement this method in a practical pulsar
timing programme?  As we have discussed, a \textit{sina qua non} is a
large training set of high-quality single pulses.  These can be
obtained in a single dedicated session, possibly on a
target-of-opportunity basis when a pulsar is unusually bright due to
scintillation.  Subsequent observations should record single pulses,
but possibly with reduced quality.  Compared to typical folded
sub-integrations, full-resolution single pulse data from young
(millisecond) pulsars requires about 100 (10,000) times more storage.
While full-resolution data is intrinsically interesting for studying
the pulsar mechanism, our method is based on fully reduced (in
frequency and polarization) data, and recording such archives requires
only modest additional storage.  We strongly advocate such a commensal
programme, as it will both enable our proposed study of jitter/SWIMS
and help inform the design of future pulsar timing instrumentation.

\bibliographystyle{mn2e}
\bibliography{ms_v3}

\begin{thebibliography}{40}
\expandafter\ifx\csname natexlab\endcsname\relax\def\natexlab#1{#1}\fi

\bibitem[{{Antoniadis} {et~al}\mbox{.}(2013){Antoniadis}, {Freire}, {Wex},
  {Tauris}, {Lynch}, {van Kerkwijk}, {Kramer}, {Bassa}, {Dhillon}, {Driebe},
  {Hessels}, {Kaspi}, {Kondratiev}, {Langer}, {Marsh}, {McLaughlin},
  {Pennucci}, {Ransom}, {Stairs}, {van Leeuwen}, {Verbiest}, \&
  {Whelan}}]{Antoniadis13}
{Antoniadis} J. {et~al.}, 2013, Science, 340, 448

\bibitem[{{Burke-Spolaor} {et~al}\mbox{.}(2012){Burke-Spolaor}, {Johnston},
  {Bailes}, {Bates}, {Bhat}, {Burgay}, {Champion}, {D'Amico}, {Keith},
  {Kramer}, {Levin}, {Milia}, {Possenti}, {Stappers}, \& {van
  Straten}}]{Burke-Spolaor12}
{Burke-Spolaor} S. {et~al.}, 2012, \mnras, 423, 1351

\bibitem[{{Cairns}(2004)}]{Cairns04}
{Cairns} I.~H., 2004, \apj, 610, 948

\bibitem[{{Cairns}, {Johnston} \& {Das}(2001){Cairns}, {Johnston}, \&
  {Das}}]{Cairns01}
{Cairns} I.~H., {Johnston} S., {Das} P., 2001, \apjl, 563, L65

\bibitem[{{Cognard} {et~al}\mbox{.}(1996){Cognard}, {Shrauner}, {Taylor}, \&
  {Thorsett}}]{Cognard96}
{Cognard} I., {Shrauner} J.~A., {Taylor} J.~H., {Thorsett} S.~E., 1996, \apjl,
  457, L81

\bibitem[{{Cordes}(1975)}]{Cordes75}
{Cordes} J.~M., 1975, PhD thesis, California Univ., San Diego.

\bibitem[{{Demorest} {et~al}\mbox{.}(2013){Demorest}, {Ferdman}, {Gonzalez},
  {Nice}, {Ransom}, {Stairs}, {Arzoumanian}, {Brazier}, {Burke-Spolaor},
  {Chamberlin}, {Cordes}, {Ellis}, {Finn}, {Freire}, {Giampanis}, {Jenet},
  {Kaspi}, {Lazio}, {Lommen}, {McLaughlin}, {Palliyaguru}, {Perrodin},
  {Shannon}, {Siemens}, {Stinebring}, {Swiggum}, \& {Zhu}}]{Demorest13}
{Demorest} P.~B. {et~al.}, 2013, \apj, 762, 94

\bibitem[{{Demorest} {et~al}\mbox{.}(2010){Demorest}, {Pennucci}, {Ransom},
  {Roberts}, \& {Hessels}}]{Demorest10}
{Demorest} P.~B., {Pennucci} T., {Ransom} S.~M., {Roberts} M.~S.~E., {Hessels}
  J.~W.~T., 2010, \nat, 467, 1081

\bibitem[{{Granet} {et~al}\mbox{.}(2005){Granet}, {Zhang}, {Forsyth}, {Graves},
  {Doherty}, {Greene}, {James}, {Sykes}, {Bird}, {Sinclair}, {Moorey}, \&
  {Manchester}}]{Granet05}
{Granet} C. {et~al.}, 2005, IEEE Antennas Propagation Magazine, 47, 13

\bibitem[{{Gwinn} \& {Johnson}(2011)}]{Gwinn11}
{Gwinn} C.~R., {Johnson} M.~D., 2011, \apj, 733, 51

\bibitem[{{Hankins} {et~al}\mbox{.}(2003){Hankins}, {Kern}, {Weatherall}, \&
  {Eilek}}]{Hankins03}
{Hankins} T.~H., {Kern} J.~S., {Weatherall} J.~C., {Eilek} J.~A., 2003, \nat,
  422, 141

\bibitem[{{Hermsen} {et~al}\mbox{.}(2013){Hermsen}, {Hessels}, {Kuiper}, {van
  Leeuwen}, {Mitra}, {de Plaa}, {Rankin}, {Stappers}, {Wright}, {Basu},
  {Alexov}, \& et~al.}]{Hermsen13}
{Hermsen} W. {et~al.}, 2013, Science, 339, 436

\bibitem[{{Hewish} {et~al}\mbox{.}(1968){Hewish}, {Bell}, {Pilkington},
  {Scott}, \& {Collins}}]{Hewish68}
{Hewish} A., {Bell} S.~J., {Pilkington} J.~D.~H., {Scott} P.~F., {Collins}
  R.~A., 1968, \nat, 217, 709

\bibitem[{{Hobbs} {et~al}\mbox{.}(2014){Hobbs}, {Dai}, {Manchester}, {Shannon},
  {Kerr}, {Lee}, \& {Xu}}]{Hobbs14}
{Hobbs} G., {Dai} S., {Manchester} R.~N., {Shannon} R.~M., {Kerr} M., {Lee}
  K.~J., {Xu} R., 2014, ArXiv e-prints

\bibitem[{{Hobbs}, {Lyne} \& {Kramer}(2010){Hobbs}, {Lyne}, \&
  {Kramer}}]{Hobbs10b}
{Hobbs} G., {Lyne} A.~G., {Kramer} M., 2010, \mnras, 402, 1027

\bibitem[{{Jenet}, {Anderson} \& {Prince}(2001){Jenet}, {Anderson}, \&
  {Prince}}]{Jenet01}
{Jenet} F.~A., {Anderson} S.~B., {Prince} T.~A., 2001, \apj, 546, 394

\bibitem[{{Johnston} {et~al}\mbox{.}(2001){Johnston}, {van Straten}, {Kramer},
  \& {Bailes}}]{Johnston01}
{Johnston} S., {van Straten} W., {Kramer} M., {Bailes} M., 2001, \apjl, 549,
  L101

\bibitem[{{Kaspi}, {Taylor} \& {Ryba}(1994){Kaspi}, {Taylor}, \&
  {Ryba}}]{Kaspi94}
{Kaspi} V.~M., {Taylor} J.~H., {Ryba} M.~F., 1994, \apj, 428, 713

\bibitem[{{Keith} {et~al}\mbox{.}(2013){Keith}, {Coles}, {Shannon}, {Hobbs},
  {Manchester}, {Bailes}, {Bhat}, {Burke-Spolaor}, {Champion}, {Chaudhary},
  {Hotan}, {Khoo}, {Kocz}, {Os{\l}owski}, {Ravi}, {Reynolds}, {Sarkissian},
  {van Straten}, \& {Yardley}}]{Keith13}
{Keith} M.~J. {et~al.}, 2013, \mnras, 429, 2161

\bibitem[{{Kramer} {et~al}\mbox{.}(2006){Kramer}, {Lyne}, {O'Brien}, {Jordan},
  \& {Lorimer}}]{Kramer06}
{Kramer} M., {Lyne} A.~G., {O'Brien} J.~T., {Jordan} C.~A., {Lorimer} D.~R.,
  2006, Science, 312, 549

\bibitem[{{Kramer} \& {Wex}(2009)}]{Kramer09}
{Kramer} M., {Wex} N., 2009, Classical and Quantum Gravity, 26, 073001

\bibitem[{{Krishnamohan} \& {Downs}(1983)}]{Krishnamohan83}
{Krishnamohan} S., {Downs} G.~S., 1983, \apj, 265, 372

\bibitem[{{Kulkarni}(1989)}]{Kulkarni89}
{Kulkarni} S.~R., 1989, \aj, 98, 1112

\bibitem[{{Lattimer} \& {Prakash}(2004)}]{Lattimer04}
{Lattimer} J.~M., {Prakash} M., 2004, Science, 304, 536

\bibitem[{{Lyne} {et~al}\mbox{.}(2010){Lyne}, {Hobbs}, {Kramer}, {Stairs}, \&
  {Stappers}}]{Lyne10}
{Lyne} A., {Hobbs} G., {Kramer} M., {Stairs} I., {Stappers} B., 2010, Science,
  329, 408

\bibitem[{{Manchester} {et~al}\mbox{.}(2013){Manchester}, {Hobbs}, {Bailes},
  {Coles}, {van Straten}, {Keith}, {Shannon}, {Bhat}, {Brown}, {Burke-Spolaor},
  {Champion}, {Chaudhary}, {Edwards}, {Hampson}, {Hotan}, {Jameson}, {Jenet},
  {Kesteven}, {Khoo}, {Kocz}, {Maciesiak}, {Oslowski}, {Ravi}, {Reynolds},
  {Sarkissian}, {Verbiest}, {Wen}, {Wilson}, {Yardley}, {Yan}, \&
  {You}}]{Manchester13}
{Manchester} R.~N. {et~al.}, 2013, \pasa, 30, 17

\bibitem[{{Os{\l}owski} {et~al}\mbox{.}(2014){Os{\l}owski}, {van Straten},
  {Bailes}, {Jameson}, \& {Hobbs}}]{Oslowski14}
{Os{\l}owski} S., {van Straten} W., {Bailes} M., {Jameson} A., {Hobbs} G.,
  2014, \mnras, 441, 3148

\bibitem[{{Os{\l}owski} {et~al}\mbox{.}(2013){Os{\l}owski}, {van Straten},
  {Demorest}, \& {Bailes}}]{Oslowski13}
{Os{\l}owski} S., {van Straten} W., {Demorest} P., {Bailes} M., 2013, \mnras,
  430, 416

\bibitem[{{Os{\l}owski} {et~al}\mbox{.}(2011){Os{\l}owski}, {van Straten},
  {Hobbs}, {Bailes}, \& {Demorest}}]{Oslowski11}
{Os{\l}owski} S., {van Straten} W., {Hobbs} G.~B., {Bailes} M., {Demorest} P.,
  2011, \mnras, 418, 1258

\bibitem[{{Ransom} {et~al}\mbox{.}(2014){Ransom}, {Stairs}, {Archibald},
  {Hessels}, {Kaplan}, {van Kerkwijk}, {Boyles}, {Deller}, {Chatterjee},
  {Schechtman-Rook}, {Berndsen}, {Lynch}, {Lorimer}, {Karako-Argaman}, {Kaspi},
  {Kondratiev}, {McLaughlin}, {van Leeuwen}, {Rosen}, {Roberts}, \&
  {Stovall}}]{Ransom14}
{Ransom} S.~M. {et~al.}, 2014, \nat, 505, 520

\bibitem[{{Rickett}(1975)}]{Rickett75}
{Rickett} B.~J., 1975, \apj, 197, 185

\bibitem[{{Shannon} \& {Cordes}(2010)}]{Shannon10}
{Shannon} R.~M., {Cordes} J.~M., 2010, \apj, 725, 1607

\bibitem[{{Shannon}, {Os{\l}owski} \& et~al.(2014){Shannon}, {Os{\l}owski}, \&
  et~al.}]{Shannon14}
{Shannon} R.~M., {Os{\l}owski}, {Dai} S., et~al., 2014, ArXiv e-prints

\bibitem[{{Shannon} {et~al}\mbox{.}(2013){Shannon}, {Ravi}, {Coles}, {Hobbs},
  {Keith}, {Manchester}, {Wyithe}, {Bailes}, {Bhat}, {Burke-Spolaor}, {Khoo},
  {Levin}, {Oslowski}, {Sarkissian}, {van Straten}, {Verbiest}, \&
  {Want}}]{Shannon13}
{Shannon} R.~M. {et~al.}, 2013, Science, 342, 334

\bibitem[{{Taylor}(1992)}]{Taylor92}
{Taylor} J.~H., 1992, Royal Society of London Philosophical Transactions Series
  A, 341, 117

\bibitem[{{Taylor}, {Fowler} \& {McCulloch}(1979){Taylor}, {Fowler}, \&
  {McCulloch}}]{Taylor79}
{Taylor} J.~H., {Fowler} L.~A., {McCulloch} P.~M., 1979, \nat, 277, 437

\bibitem[{{van Straten}(2004)}]{vanStraten04}
{van Straten} W., 2004, \apjs, 152, 129

\bibitem[{{van Straten}(2013)}]{vanStraten13}
{van Straten} W., 2013, \apjs, 204, 13

\bibitem[{{van Straten} \& {Bailes}(2011)}]{vanStraten11}
{van Straten} W., {Bailes} M., 2011, \pasa, 28, 1

\bibitem[{{Yardley} {et~al}\mbox{.}(2011){Yardley}, {Coles}, {Hobbs},
  {Verbiest}, {Manchester}, {van Straten}, {Jenet}, {Bailes}, {Bhat},
  {Burke-Spolaor}, {Champion}, {Hotan}, {Oslowski}, {Reynolds}, \&
  {Sarkissian}}]{Yardley11}
{Yardley} D.~R.~B. {et~al.}, 2011, \mnras, 414, 1777

\end{thebibliography}

\section*{Acknowledgments}

We are grateful to Stefan {Os{\l}owski} for input on an early draft
and to the anonymous referee for helpful criticism that improved the
paper.  The Parkes radio telescope is part of the Australia Telescope,
which is funded by the Commonwealth Government for operation as a
National Facility managed by CSIRO.

\end{document}